\documentclass[twocolumn,aps,prc,superscriptaddress,showpacs]{revtex4}

\usepackage{graphicx}
\draft
\usepackage{color}

\begin{document}


\title{Collision system size dependence of dihadron azimuthal correlations in
 ultra-relativistic heavy ion collisions}

 \author{ S. Zhang}
\affiliation{Shanghai Institute of Applied Physics, Chinese
Academy of Sciences, Shanghai 201800, China}
\author{ Y. H. Zhu}
\affiliation{Shanghai Institute of Applied Physics, Chinese
Academy of Sciences, Shanghai 201800, China} \affiliation{Graduate School of
  the Chinese Academy of Sciences, Beijing 100080, China}
\author{ G. L. Ma}
\affiliation{Shanghai Institute of Applied Physics, Chinese
Academy of Sciences, Shanghai 201800, China}
\author{ Y. G. Ma\footnote{Corresponding author: ygma@sinap.ac.cn}}
\affiliation{Shanghai Institute of Applied Physics, Chinese
Academy of Sciences, Shanghai 201800, China}
\author{ X. Z. Cai}
\affiliation{Shanghai Institute of Applied Physics, Chinese
Academy of Sciences, Shanghai 201800, China}
\author{ J. H. Chen}
\affiliation{Shanghai Institute of Applied Physics, Chinese
Academy of Sciences, Shanghai 201800, China}
\author{ C. Zhong}
\affiliation{Shanghai Institute of Applied Physics, Chinese
Academy of Sciences, Shanghai 201800, China}

\date{ \today}

\begin{abstract}
The system size dependence of dihadron azimuthal correlations in
ultra-relativistic heavy ion collision is simulated by a
multi-phase transport model. The structure of correlation
functions and yields of associated particles show clear
participant path-length dependences in collision systems with a
partonic phase. The splitting parameter ($D$) and Root Mean Square
Width ($\Delta \phi_{rms}$) of away side correlation functions
increase with collision system size from $^{14}$N+$^{14}$N to
$^{197}$Au+$^{197}$Au collisions. The double-peak structure of
away side correlation functions can only be formed in sufficient
``large'' collision systems. These properties provide some hints
to study onset of deconfinement, which is related to the QCD phase
boundary and QCD critical point, by an energy-size scan.

\end{abstract}


\maketitle

\begin{figure*}[htbp]
\resizebox{15cm}{!}{\includegraphics{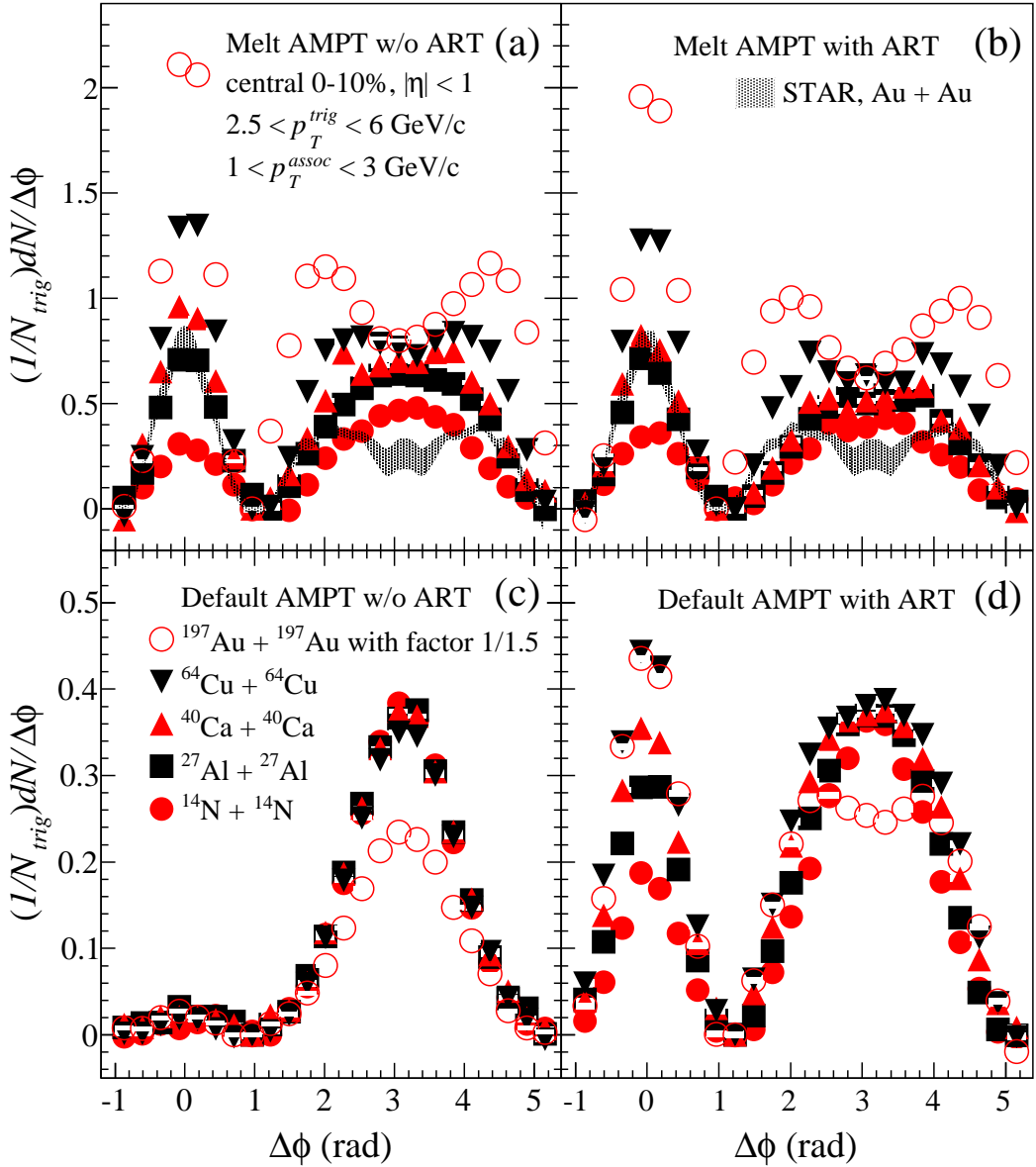}}
\caption{(Color online) Dihadron azimuthal correlation functions
for different collision systems in centrality 0-10\% at
$\sqrt{s_{NN}}$ = 200  GeV; the shadowing area from the STAR
data~\cite{sideward-peak1}.}
\label{corr-fun} 
\end{figure*}

\section{Introduction}
Quantum Chromodynamics (QCD) calculation predicted an exotic
quark-gluon matter in QCD phase diagram~\cite{QCD-phasetran} may
be created  in the early stage of heavy ion collisions at
ultra-relativistic energy~\cite{White-papers}. For mapping the QCD
phase diagram and locating QCD phase boundary and critical point
~\cite{phase-th}, one needs to find a way to vary temperature,
$T$, and chemical potential, $\mu_{B}$.  The NA61 collaboration
and NA49-future collaboration~\cite{NA61EA} suggested that it can
be achieved via a systematic energy (E) and system size (A)
($E-A$) scan. The present work focuses on the latter in different
collision systems with partonic phase or hadron gas at
$\sqrt{s_{NN}}$ = 200 GeV.

Jet quenching phenomenon has been theoretically predicted
~\cite{jet-quenching} and experimentally observed~\cite{jet-ex}.
So far, dihadron azimuthal correlation has been demonstrated as a
good method to reconstruct particle and energy distribution
induced by the quenched jet. In experiment,
 a double-peak structure was found on the away side  of dihadron azimuthal
 correlation functions~\cite{sideward-peak1,sideward-peak2,sideward-peak3}
 and the indication of conical emission of charged hadrons was reported by
 the STAR collaboration~\cite{sideward-peak4}. A significant centrality
 dependence of double-peak structure of away-side  correlation functions
 was observed by the PHENIX collaboration~\cite{sideward-peak3} and
 theoretically simulated in Ref.~\cite{di-hadron}. The centrality
 dependence results indicate that the structure of away-side  correlation
 function is sensitive to system size of the reaction zone.

These interesting phenomena provoke some theorists into explaining
the physical mechanisms for origins of the double-peak structure,
such as Cherenkov-like gluon radiation model~\cite{Koch},
medium-induced gluon bremsstrahlung
radiation~\cite{large-angle,opaque-media-radiation}, shock wave
model in hydrodynamic equations~\cite{Casalderrey}, waking the
colored plasma and sonic Mach cones~\cite{Ruppert}, sonic booms
and diffusion wakes in thermal gauge-string
duality~\cite{sonic-booms}, jet deflection~\cite{deflection} and
strong parton cascade mechanism
~\cite{di-hadron,three-hadron,time-evolution,pt-dependence,glma-sqm09}.

All of the experimental and theoretical works suggest that the
double-peak  phenomenon is a good probe to study the hot and dense
medium created in ultra-relativistic  heavy ion
collisions~\cite{hard-hard-ex,soft-soft-ex}. In this paper we
investigate the double-peak structure in different collision
systems, namely $^{14}$N + $^{14}$N, $^{16}$O + $^{16}$O,
$^{23}$Na + $^{23}$Na, $^{27}$Al + $^{27}$Al, $^{40}$Ca +
$^{40}$Ca, $^{64}$Cu + $^{64}$Cu and $^{197}$Au + $^{197}$Au at
$\sqrt{s_{NN}}$ = 200 GeV. The collision system size dependence of
the double-peak structure of away-side   correlation function is
focused.  We present participant path-length $\nu$ =
2$N_{bin}$/$N_{part}$~\cite{nu-define} ($N_{bin}$ and $N_{part}$
are the number of binary collision and participants, respectively)
dependence of the double-peak structure of away-side  correlation
function in the most central collisions (0-10\%). In addition, the
behavior of the away-side  correlation function is also discussed
at given various participant path-length $\nu$ $\approx$ 1.90 and
2.28. From these results, one can see that  the structure of
away-side  correlation function changes near $^{40}$Ca + $^{40}$Ca
collisions at $\sqrt{s_{NN}}$ = 200 GeV in central collisions
(0-10\%), but there is no significant system dependence of the
correlation at given participant path-length $\nu$.  These results
show significant degree of freedom dependence in the system with a
partonic phase or with a pure hadron gas \cite{Song}, which
provides potential information about the onset of deconfinement.
This method supports the $E-A$ scan in experiment to investigate
the QCD phase transition boundary and QCD critical point.

The paper is organized as follows. The following section
introduces  the model as well as the analysis method for dihadron
azimuthal correlations. Section III describes the detailed results
and discussions, which includes the structure of dihadron
azimuthal  correlation functions, yields and mean transverse
momentum of associated particles, the splitting parameter ($D$)
and $\Delta \phi_{rms}$ dispersion of away-side correlation
functions and the corresponding results at given participant
path-lengths. The nuclear modification factor, $R_{CP}$, are also
presented in Section III. Section IV gives a summary.

\section{Model and analysis method}

In this paper, a multi-phase transport model (AMPT)~\cite{AMPT},
which  is a hybrid model, is employed to study dihadron azimuthal
correlations. It includes four main components to describe the
physics in relativistic heavy ion collisions: 1) the initial
conditions from HIJING model~\cite{HIJING}, 2) partonic
interactions modeled by a Parton Cascade model (ZPC)~\cite{ZPC},
3) hadronization (discussed later), 4) hadronic rescattering
simulated by A Relativistic Transport (ART) model~\cite{ART}.
Excited strings from HIJING are melted into partons in the AMPT
version with string melting mechanism~\cite{SAMPT} (abbr.
{`\it{the Melt AMPT version}'}) and a simple quark coalescence
model is used to combine the partons into hadrons. In the default
version of AMPT model~\cite{DAMPT}(abbr. {`\it{the Default AMPT
version}'}), minijet partons are recombined with their parent
strings when they stop interactions and the resulting strings are
converted to hadrons via the Lund string fragmentation
model~\cite{Lund}. The Melt AMPT version undergoes a partonic
phase, while a pure hadron gas is in the Default AMPT version.
Details of the AMPT model can be found in a review
paper~\cite{AMPT} and previous works~\cite{AMPT,SAMPT,Jinhui}.

The analysis method for dihadron azimuthal correlations is similar
to that used in previous experiments
~\cite{soft-soft-ex,sideward-peak2}, which describes the azimuthal
correlation between a high $p_{T}$ particle (trigger particle) and
low $p_{T}$ particles (associated particles). The raw signal can
be obtained by accumulating pairs of trigger and associated
particles into $\Delta\phi = \phi_{assoc} - \phi_{trig}$
distributions in the same event. The background which is expected
mainly from elliptic flow is simulated by mixing event
method~\cite{soft-soft-ex,sideward-peak2}. By selecting the
associated particles in different events whose centralities are
every close to that of the same event (raw signals), the
$\Delta\phi$ distribution can be obtained as the corresponding
background. The background is subtracted from raw signal by using
A Zero Yield At Minimum (ZYAM) assumption as that used in
experimental analysis~\cite{sideward-peak2} (See our detailed
analysis in Ref.~\cite{di-hadron}).

\section{Results and discussions}

\subsection{Structure of dihadron correlation function}

\begin{table*}[htb]
\begin{center}
\caption{ $N_{part}(CSYS)$, $N_{bin}(CSYS)$, $\nu(CSYS)$ =
  $\frac{2N_{bin}(CSYS)}{N_{part}(CSYS)}$, $n_{col}^{parton}$ in different
    collision system at $\sqrt{s_{NN}}$ = 200 GeV for centrality 0-10 \%, the
    value in blanket is taken from Glauber Model~\cite{sys-glauber}.}
\begin{tabular}{|c|c|c|c|c|c|c|c|}\hline
$CSYS$ & $^{14}$N + $^{14}$N & $^{16}$O + $^{16}$O & $^{23}$Na +
$^{23}$Na & $^{27}$Al + $^{27}$Al & $^{40}$Ca + $^{40}$Ca &
$^{64}$Cu + $^{64}$Cu & $^{197}$Au + $^{197}$Au \\\hline
$N_{part}(CSYS)$ & 20.78 & 24.25 & 35.92 & 43.61 & 65.97 & 107.04
(99.0) & 343.32 (325.9)\\\hline $N_{bin}(CSYS)$ & 19.63 & 23.69
& 41.01 & 54.34 & 91.15 & 179.98 (188.8) & 914.71
(939.4)\\\hline $\nu(CSYS)$ =
$\frac{2N_{bin}(CSYS)}{N_{part}(CSYS)}$ & 1.89 & 1.95 & 2.28 &
2.49 & 2.76 & 3.36 (3.8) & 5.33 (5.7)\\\hline $n_{col}^{parton}$
& 1.31 & 1.44 & 1.93 & 2.23 & 2.79 & 3.80 & 7.24
\\\hline
\end{tabular}

\label{talble-samecentral}
\end{center}
\end{table*}

The participant path-length, defined as $\nu$ =
2$N_{bin}$/$N_{part}$~\cite{nu-define},  can describe degree of
multiple collisions between participants in the early stage of
heavy ion collisions and characterize the size of the reaction
zone. The $n_{col}^{parton}$  represents average collision number
of partons in the Melt AMPT version. The values of $\nu$ and
$n_{col}^{parton}$ significantly increase with varying collision
system ($CSYS$) from ``small'' size to ``large'' size at
$\sqrt{s_{NN}}$ = 200 GeV in the most central collisions (0-10\%)
as shown in Table~\ref{talble-samecentral}. From this table, we
can see the multiple collisions are more enhanced in ``large''
size collision system than in ``small'' size one. The values of
$N_{part}$, $N_{bin}$ and $\nu$ are comparable to those from the
Glauber Model~\cite{sys-glauber} for both $^{64}$Cu + $^{64}$Cu
and $^{197}$Au + $^{197}$Au collisions.

\begin{figure}[htbp]
\resizebox{8.6cm}{!}{\includegraphics{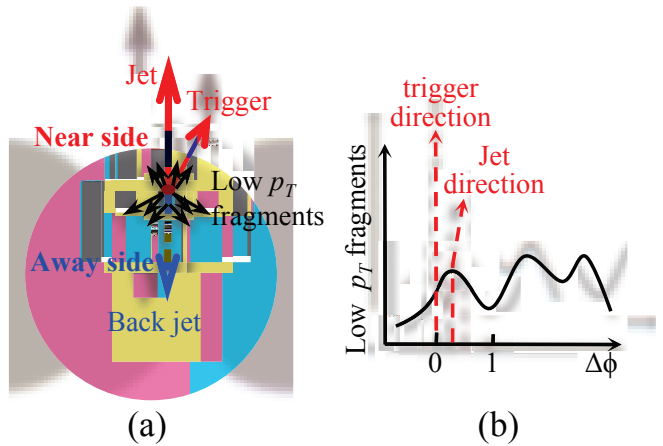}} 
\caption{(Color online) Schematic illustrations, (a) fragments from jet in relativistic heavy ion collisions, (b) azimuthal correlations between trigger particle and low $p_T$ fragments.}
\label{frag-jet} 
\end{figure}

Figure~\ref{corr-fun} shows dihadron azimuthal correlation
functions of different  collision systems in the most central
(0-10\%) collisions at $\sqrt{s_{NN}}$ = 200 GeV. The correlation
functions are calculated in the kinetic windows,  1 $<
p_{T}^{assoc} <$ 3 GeV/$c$ as well as 2.5 $< p_{T}^{trig} <$ 6
GeV/$c$ and $|\eta| <$ 1. It shows that the structure of away-side
correlation function changes from the Gaussian-like distribution
to double-peak structure, obviously emerging in $^{40}$Ca +
$^{40}$Ca collisions, with varying collision system from $^{14}$N
+ $^{14}$N to $^{197}$Au + $^{197}$Au collisions in the Melt AMPT
version. In this figure, the amplitude of the correlation function
becomes higher with the increasing of collision system size. The
associated particles in the Melt AMPT version are more abundant
than those in the Default AMPT version.

The yields of associated particles on near-side correlation
functions from the Default  AMPT version without hadronic
rescattering is very small. The jet pair from HIJING~\cite{HIJING}
keeps momentum conservation law and number of particles from jet
fragmentation dose not balance on the near side and away side,
which was investigated in p + p collisions~\cite{KFXin-CPL}. In
the Default AMPT version without hadronic rescattering, the
Gaussian-shaped fragments around the jet direction are emitted
from the jet, while the trigger particle in the fragments are not
always along the direction of the jet, a schematic illustration
shown in panel (a) of Fig~\ref{frag-jet}. The azimuthal
correlations between trigger particle and low $p_T$ fragments can
be obtained in panel (b) in Fig~\ref{frag-jet}. The near-side peak
are around direction of near-side jet and shifts from direction of
trigger particle in some cases. As an integral result, the
amplitude of near side will be very small in the Default AMPT
version without hadronic rescattering.

It is obvious that hadronic rescattering enhances the correlation
on the near side  and broadens the structure on the away side  in
the Default AMPT version. As we have already demonstrated, the
structure of dihadron correlation functions is more reasonable in
the Melt AMPT version than in the Default AMPT
version~\cite{di-hadron}. However the Default AMPT version is used
to compare the properties of the double-peak structure in partonic
phase and in hadron gas, especially for ``small'' size collision
system.  For investigating the properties of collision system size
dependences of away-side  dihadron azimuthal correlations, we
extract the associated particle yield $N_{away}^{assoc}$,
splitting parameter ($D$) (half distance between double peaks on
the away side ), Root Mean Square Width ($\Delta \phi_{rms}$) and
mean transverse momentum $\langle p_{T}\rangle_{away}^{assoc}$ of
away-side  associated particles, which will be discussed later,
respectively.

\subsection{Yield of associated particles}
\begin{figure}[htbp]
\resizebox{8.6cm}{!}{\includegraphics{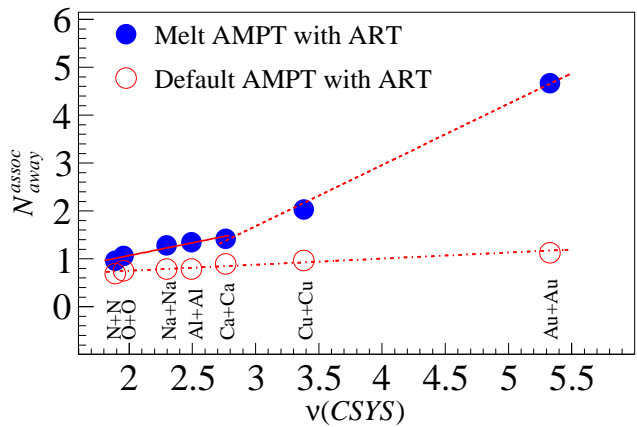}} 
\caption{(Color online) Yield of associated particles on away-side
correlation functions, $N_{away}^{assoc}$, as a function of
$\nu(CSYS)$ by the Melt/Default AMPT version in centrality 0-10\%
at $\sqrt{s_{NN}}$ = 200 GeV.}
\label{yield} 
\end{figure}

The $\nu(CSYS)$ dependence of $N_{away}^{assoc}$ is shown in
Fig.~\ref{yield}  from the Melt/Default AMPT version,
respectively. It presents a significant increasing trend of
$N_{away}^{assoc}$ with varying the collision system from $^{14}$N
+ $^{14}$N to $^{197}$Au + $^{197}$Au collisions in the most
central collisions (0-10\%) at $\sqrt{s_{NN}}$ = 200 GeV in the
Melt AMPT version. The Default AMPT version, with a hadronic gas,
does not result in a rapidly increasing dependence trend.  In the
Melt AMPT version, the dependence trend indicates the jet
correlation information can be inherited by more particles in a
partonic phase than in a hadronic gas, especially in ``large''
size collision system. Furthermore it implies that the interaction
strength in ``large'' size collision system is more significant
than that in ``small'' size collision system, and while strong
parton cascade plays a dominant role in dihadron azimuthal
correlations in the Melt AMPT version.  It is interesting that the
increasing slope  of $N_{away}^{assoc}$ vs $\nu(CSYS)$  from the
linear fitting in the Melt AMPT version is quicker after $^{40}Ca$
+ $^{40}Ca$ collision system, where clear double-peak structure
emerges, than that in small systems. This property indicates the
double-peak (Mach-like) structure can enhance associated particles
yields of jet correlations partially.

\subsection{$\Delta \phi_{rms}$ and splitting parameter on the away side }
\begin{figure}[htbp]
\resizebox{8.6cm}{!}{\includegraphics{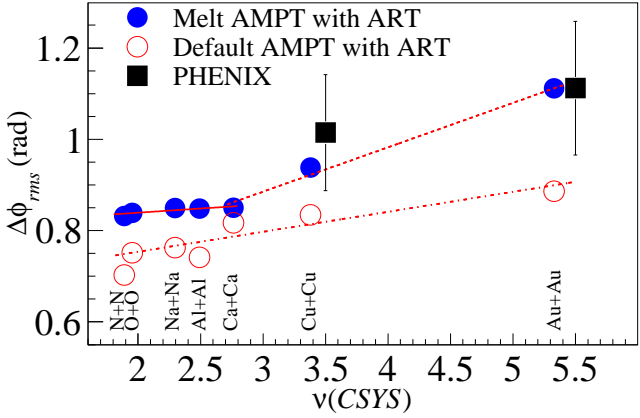}} 
\caption{(Color online) $\Delta \phi_{rms}$  of the away-side
correlation functions as a function of $\nu(CSYS)$ in the
Melt/Defualt AMPT version for the centrality 0-10\% at
$\sqrt{s_{NN}}$ = 200 GeV; Square from PHENIX
data~\cite{sideward-peak2,rmsdata}.}
\label{rms} 
\end{figure}

\begin{figure}[htbp]
\resizebox{8.6cm}{!}{\includegraphics{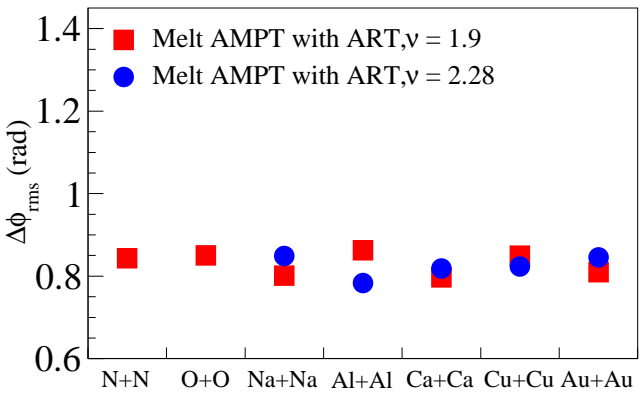}}
\caption{(Color online) $\Delta \phi_{rms}$ on away-side  of the
correlation functions at given $\nu(CSYS)$ $\approx$ 1.90, 2.28 at
$\sqrt{s_{NN}}$ = 200 GeV.} \label{rms-samenu}
\end{figure}

Root Mean Square Width ($\Delta \phi_{rms}$) of away-side
correlation function is defined as
\begin{displaymath}
\Delta
\phi_{rms}=\sqrt{\frac{\sum\limits_{away}{(\Delta\phi-\Delta\phi_{m})^{2}(1/N_{trig})(dN/d\Delta\phi)}}{\sum\limits_{away}{(1/N_{trig})(dN/d\Delta\phi)}}},
\end{displaymath}
where $\Delta\phi_{m}$ is the mean $\Delta\phi$ of away-side
correlation function and  it approximates to $\pi$. $\Delta
\phi_{rms}$ can describe broadening of away-side  correlation
function relative to $\pi$ which is direction of back jet. By
$\Delta \phi_{rms}$, one can investigate diffusion degree of the
associated particles relative to back jet. The $\nu(CSYS)$
dependences of $\Delta \phi_{rms}$ in the Melt/Default AMPT
version are shown in Fig.~\ref{rms}, respectively.  $\Delta
\phi_{rms}$ from the Melt AMPT version are consistent with PHENIX
data~\cite{sideward-peak2,rmsdata} in Cu + Cu and Au + Au
collisions. $\Delta\phi_{rms}$ increases from $^{14}$N + $^{14}$N
collisions to $^{197}$Au + $^{197}$Au collisions in the Melt AMPT
version and the increasing trend is not so quick in the Default
AMPT version. The increasing trend of $\Delta \phi_{rms}$ shows
broadening of away-side correlation functions with increasing size
of collision system. It indicates that the jet correlation
information can reach faraway relative to direction of jet with
changing the collision system from ``small'' size one to ``large''
size in a partonic phase. It is remarkable that the increasing
trend of $\Delta \phi_{rms}$  from the linear fitting in the Melt
AMPT version shows two different slope after and before  $^{40}$Ca
+ $^{40}$Ca collision system, where clear double-peak structure
emerges. This interesting result indicates that the double-peak
(Mach-like) phenomenon from quenched-jet pushes evolution of jet
correlations. At given $\nu(CSYS)$ case, we investigate system
dependence of the double-peak structure in the Melt AMPT version.
The $\Delta \phi_{rms}$  keeps a flat pattern with varying
collision system in Fig.~\ref{rms-samenu} at given $\nu(CSYS)$
$\approx$ 1.9 and 2.28. The $\Delta \phi_{rms}$ independence of
collision system at given $\nu(CSYS)$ implies that the broadening
of the away-side correlation function is only sensitive to how
violent interactions in collision system, which determines how far
the jet correlation information spreads. These results suggest
that the back jet modification in the medium created in heavy ion
collisions with a partonic phase is more obvious in the ``large''
size collision system than in ``small'' size one.

\begin{figure}[htbp]
\resizebox{8.6cm}{!}{\includegraphics{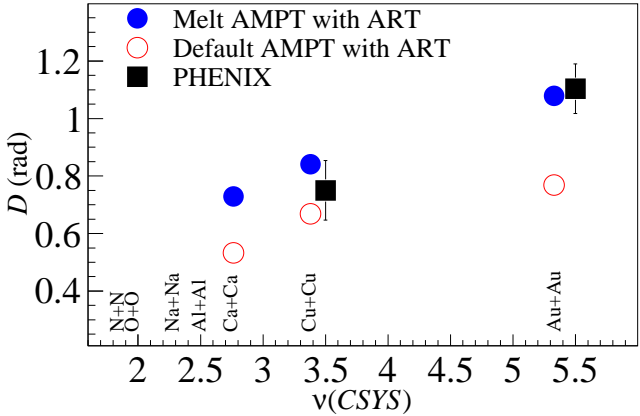}} 
\caption{(Color online) Splitting parameter ($D$) of the away-side
correlation functions as a function of $\nu(CSYS)$ in the
Melt/Defualt AMPT version for the centrality 0-10\% at
$\sqrt{s_{NN}}$ = 200 GeV; Square from PHENIX
data~\cite{sideward-peak3}.}
\label{splitd} 
\end{figure}

\begin{figure}[htbp]
\resizebox{8.6cm}{!}{\includegraphics{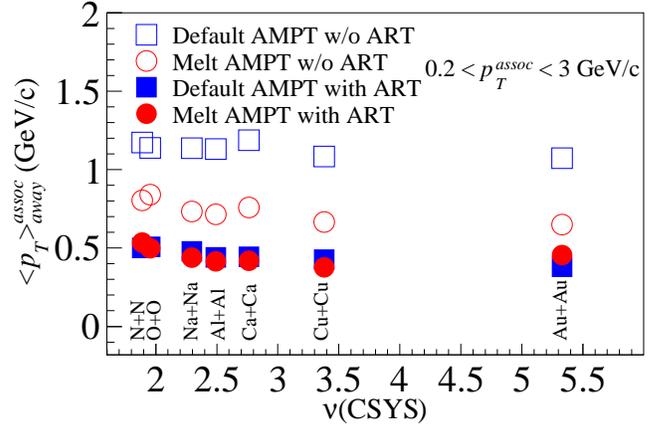}} 
 \caption{(Color online) Mean transverse momentum
 ($\langle p_{T}\rangle_{away}^{assoc}$) of the away-side  correlation
 functions as a function of $\nu(CSYS)$ in the Melt/Default AMPT version
 in centrality 0-10\% at $\sqrt{s_{NN}}$ = 200 GeV. 
 }
\label{meanpT-nu} 
\end{figure}

\begin{figure*}[htbp]
\resizebox{17cm}{!}{\includegraphics{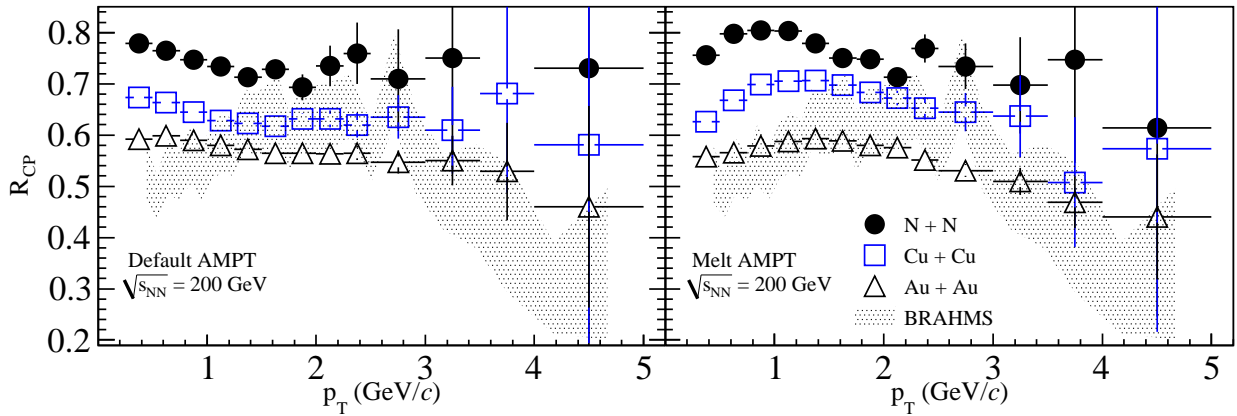}} 
 \caption{(Color online) $R_{CP}$ in the Default AMPT Version at $\sqrt{s_{NN}}$ = 200 GeV, the shadowing from BRAHMS data~\cite{RCP-BRAHMS} in Au + Au collisions.}
\label{RCP-AMPT}
\end{figure*}

The splitting parameter ($D$) is another useful observable to
characterize the structure of  the double-peak of away-side
correlation function, and further discloses essential of jet
modification. The $\nu(CSYS)$ dependence of splitting parameter
($D$) in the Melt/Default AMPT version are shown in
Fig.~\ref{splitd}, respectively, in the most central collisions at
$\sqrt{s_{NN}}$ = 200 GeV. Double Gaussian function is not used to
fit away-side  correlation functions since there is no obvious
double-peak structure from in $^{14}$N + $^{14}$N to in $^{27}$Al
+ $^{27}$Al collisions. The splitting parameter ($D$) increases
from ``small'' size collision system to ``large'' size one in the
Melt AMPT version. It is obvious that the splitting parameter
($D$) is smaller in the Default AMPT version than in the Melt AMPT
version. The results in the Melt AMPT version are comparable to
PHENIX data~\cite{sideward-peak3} in Cu + Cu and Au + Au
collisions due to effect of parton cascade in the Melt AMPT
version~\cite{di-hadron}. The parton interaction cross section is
taken to be 10 mb, which is reasonable for reproducing elliptic
flow and dihadron azimuthal correlations in the Melt AMPT
version~\cite{AMPT,SAMPT,Jinhui,di-hadron}. In addition,
Reference~\cite{jet-ex} demonstrated that there is a strong
suppression of inclusive yield and back-to-back correlations at
high $p_{T}$ in $^{197}$Au + $^{197}$Au collisions relative to
that in d + $^{197}$Au collisions at $\sqrt{s_{NN}}$ = 200 GeV.
Theoretical studies suggest that the emission angle relative to
jet should be about 1.23 rad for QGP, 1.11 rad for hadronic gas
and zero for mixed phase~\cite{Casalderrey}. The radiation
mechanism for the double-peak structure~\cite{Koch} suggests the
more energetic the jet, the smaller the emission angle. And in
this calculation it indicates the double-peak structure is more
obvious in a partonic phase than in a hadron gas. A quick change
of the double-peak structure is expected while one experimentally
scan the system size or beam energy near the QCD critical point or
phase transition boundary. It implies that the double-peak
structure and jet modification is sensitive to the effective
degree of freedom of  the dense medium created in heavy ion
collisions, which is related to onset of deconfinement.

From these results, it can be concluded that a considerable
``large'' collision  system is necessary and the strong parton
cascade is important for forming the double-peak structure of
away-side  correlation function. An onset of the obvious
double-peak structure takes place around the collision mass range
near $^{40}$Ca + $^{40}$Ca collisions. This phenomenon indicates
that the correlation is sensitive to $\nu(CSYS)$ and
$n_{col}^{parton}$, i.e. the correlation depends on the collision
system size and the violent degree of the partonic interaction in
a partonic phase. The results in different group in a parton phase
and a pure hadron gas imply the correlations are sensitive to the
effective degree of freedom of the dense medium created in relativistic
heavy ion collisions, which can give us some hints of the onset of
deconfinement in the system-size viewpoint.

\subsection{Mean transverse momentum of associated particles in central collisions}
For further investigating the contribution of parton cascade and
hadronic rescattering  to jet modification, we calculate mean
transverse momentum of away-side  associated particles in central
collisions at $\sqrt{s_{NN}}$ = 200 GeV and $p_{T}^{assoc}$ is
fixed at (0.2, 3) GeV/$c$. $\langle p_{T} \rangle_{away}^{assoc}$
is presented as a function of $\nu(CSYS)$ in the Melt/Default AMPT
version with/without hadronic rescattering in
Fig.~\ref{meanpT-nu}, respectively. $\langle
p_{T}\rangle_{away}^{assoc}$ slightly decreases with varying
collision system from ``small'' size to ``large'' size one. $
\langle p_{T} \rangle_{away}^{assoc}$ for different systems are
higher in the Default AMPT version without hadronic rescattering
than in other case. And $ \langle p_{T} \rangle_{away}^{assoc}$
are at the bottom and no difference in the Melt AMPT/Default AMPT
version with hadronic rescattering. The results in the Melt AMPT
version without hadronic rescattering are between them. These
results are consistent with what we discussed in the introduction
to AMPT model. The initial condition for AMPT is from
HIJING~\cite{HIJING}, corresponding to the AMPT without parton
cascade and hadronic rescattering, so we get the high $\langle
p_{T} \rangle_{away}^{assoc}$ in this case. And the particles will
be softened by parton cascade and hadronic rescattering. The
condition for hadronic rescattering is sensitive to how close the
distance between hadrons is in hadronic rescattering
(ART)~\cite{ART}. In the Melt AMPT version, the system undergoes
parton cascade with expansion of the system and then partons are
combined into hadrons. However in the Default AMPT version, the
system skips the parton cascade stage and most of hadrons can
participate in hadronic rescattering. Therefore hadronic
rescattering contribution is more significant in the Default AMPT
version than in the Melt AMPT version.

\subsection{Suppression of high-$p_T$ hadrons}

Jet-quenching can be investigated by dihadron azimuthal
correlations as discussed  above. Another important phenomenon
about jet-quenching is suppression of high-$p_T$ hadrons in
relativistic heavy ion collisions. The nuclear modification
factor,$R_{cp}$, describes this phenomenon, defined as,
\begin{displaymath}
R_{CP} =
\frac{\frac{d^{2}N}{p_{T}dp_{T}d\eta}(Central)/N_{bin}(Central)}{\frac{d^{2}N}{p_{T}dp_{T}d\eta}(Peripheral)/N_{bin}(Peripheral)},
\end{displaymath}
where the central and peripheral collision centralities are,
respectively, 0-10\% and 40-60\%. Figure~\ref{RCP-AMPT} shows the
$R_{CP}$ calculated in the Melt AMPT version (right panel) and in
the Default AMPT version (left panel) in relativistic heavy ion
collisions at $\sqrt{s_{NN}}$ = 200 GeV. To clearly see the system
size dependence, we present $R_{CP}$  in $^{14}$N + $^{14}$N,
$^{64}$Cu + $^{64}$Cu and $^{197}$Au + $^{197}$Au collisions and
those in other collision systems keep the similar dependent trend.
The shadowing area in this figure is from BRAHMS
data~\cite{RCP-BRAHMS}. The hadrons in the Melt AMPT version,
which undergo a partonic phase, are suppressed at high $p_{T}$ and
those from the Default AMPT version, in a hadron gas, are not
suppressed obviously. The $R_{CP}$ in $^{197}$Au + $^{197}$Au
collisions from the Melt AMPT version is consistent with the
experimental data~\cite{RCP-BRAHMS}. We can conclude that there is
jet-quenching-like mechanism in the system with partonic phase and
jet-quenching phenomenon is more obvious in the ``large'' size
collision system than in ``small'' size one.

\section{summary}
In summary, the present paper discusses the collision system size
dependences of  dihadron azimuthal correlations at $\sqrt{s_{NN}}$
= 200 GeV by a multi-phase transport model. It presents the
properties of away-side  correlation functions, such as
$N_{assoc}^{away}$, $\Delta \phi_{rms}$, $D$ and $\langle
p_{T}^{away}\rangle$.  The onset of obvious double-peak structure
occurs near $^{40}$Ca + $^{40}$Ca collisions and away-side
correlation function becomes more and more broadening with the
increasing of collision system size.  The yields of associated
particles ($N_{assoc}^{away}$), $\Delta \phi_{rms}$ of away-side
correlation functions and splitting parameter ($D$) show
significant system size dependences. These results also present
the degree of freedom dependence, which is related to onset of
deconfinement. The present analysis may shed light on the QCD
phase boundary and critical point in experiment by the energy-size
scan.

\section*{Acknowledgements}
This work was supported in part by  the National Natural Science
Foundation of China under Grant No. 10979074,  10705043, 10705044,
10775167 and 10875159,  and the Shanghai Development Foundation
for Science and Technology under contract No. 09JC1416800, and the
Knowledge Innovation Project of the Chinese Academy of Sciences
under Grant No. KJCX2-YW-A14 and O95501P011 and the
Project-sponsored by SRF for ROCS, SEM. O819011012.



\end{document}